# Time-Resolved Diffusing Wave Spectroscopy for selected photon paths beyond 300 transport mean free paths

J.-M. Tualle*, H.L. Nghiem, M. Cheikh, D. Ettori, E. Tinet and S. Avrillier

*Laboratoire de Physique des Lasers (CNRS UMR 7538),*

*Université Paris 13, 99 avenue J-B. Clément, 93430 Villetaneuse, France*

This paper is devoted to the theoretical and experimental demonstration of the possibility to perform time-resolved diffusing wave spectroscopy: we successfully registered field fluctuations for selected photon path lengths that can overpass 300 transport mean free paths. Such a performance opens new possibilities for biomedical optics applications.

Biological tissues present a transmission windows in the near infrared range. For wavelengths near 800nm, light can travel deep inside the tissue and reveal a lot of pertinent information for medical diagnosis: this is the base of Near InfraRed Spectroscopy (NIRS), which, for example, allows an estimation of tissue oxygenation from absorption coefficients measurements[1]. Very interesting complementary information on the microscopic movements of the scatterers, such as the translation movement of the red blood cells, can also be obtained[2-6] from the correlation of the scattered field fluctuations in the frame of the so-called Diffusing Wave Spectroscopy (DWS)[7-10].

NIRS however suffers from information blurring due to scattering. The use of time-resolved detection has been proved to be an efficient tool to obtain more information concerning the tissue: it allows a separate measurement of the absorption and the reduced scattering coefficients, it is helpful in the measurement of the optical coefficients in a layered medium[11-14] and, concerning mammography, it can increase spatial resolution in trans-illumination[15,16]. However, even if a lot of improvements have been observed, setups that perform time-resolved detection are usually expensive and require complex technologies.

In previous papers we proposed[17-19] a new way to perform time-resolved measurements of the light scattered by a tissue. In addition to its lower cost, this method, based on the use of an interferometer and a wavelength modulated source, provides the great advantage of allowing time-resolved DWS. This possibility to study the correlation of the field fluctuations for a selected photon path length is very interesting and it has been shown, for acoustic waves, that it can considerably simplify the analysis of the correlation signal[20]. Such measurements were already performed in the past for photon paths that do not exceed 20 transport mean free paths[21]. In this paper we report the first results obtained for photon paths over 300 transport mean free paths, opening the possibility to perform measurements with several centimeters source-detector distances.

The principle of our method is the following (see the experimental set-up in fig 1): interferences between the back-scattered light and a reference beam are recorded. As we use a monochromatic laser source, there is no difficulty to record the resultant speckle pattern. The trick of this method is to induce a wavelength modulation of the laser source in order to simulate an incoherent source. To avoid problems with the modulation system bandwidth we use a sinusoidal modulation, and the radian frequency is $\omega(t) = \omega_0 + \Delta\Omega \cos(2\pi f t)$. The recorded interference signal can then be written, after rejection of the DC components:

$$s(t) \propto \Re_e \left\{ A_0 \cdot \widetilde{A}^*[\omega(t), t] \right\} \quad (1)$$

where $A_0$ is the complex amplitude of the reference field, and where $\widetilde{A}[\omega(t), t]$ corresponds to the complex amplitude of the scattered field. This equation infers of course that both fields are detected in the same mode, as it would be the case, for instance, if they are detected far from the source, or through a monomode fiber.

We have demonstrated in a previous paper[17] that $\widetilde{A}$ can be directly linked to the complex amplitude of the time resolved scattered field $A(\tau, t)$ of a virtual pulsed experiment, where $\tau$ is the photons time of flight. Equation (1) can therefore be rewritten:

$$s(t) \propto \int_{-\infty}^{+\infty} d\tau \, \Re_e \left\{ A_0 A^*[\tau, t] \exp[i \omega(t) \tau] \right\} \quad (2)$$

The only requirement is that the spectrum of the virtual pulsed source does not significantly vary on the scale of the wavelength modulation. One should note in (2) the use of two timescales: $t$, in the millisecond range, is the timescale of the registered signal and is related to the fluctuations of the scattered field, while $\tau$ corresponds to the photons time of flight, and is in the hundreds picosecond range. We consider that these timescales are not correlated. We stress here that our notations are somewhat different from usual notations in DWS[7-10], where the photon time of flight is not directly introduced.

In our experiment, the signal *s(t)* is multiplied by the numerically generated function[17]:

$$\text{Ref}(t,\tau) = \sin^4(2\pi f t)\exp[i\Delta\Omega\tau\cos(2\pi f t)] \tag{3}$$

and the result is integrated over half a period:

$$S_{DC,m}(\tau) = 2f \int_{m\Delta T/2}^{(m+1)\Delta T/2} s(t)\text{Ref}(t,\tau)\,dt \tag{4}$$

where $\Delta T = f^{-1}$ is the modulation period, and where *m* is indexing the half-period $[m\Delta T/2,(m+1)\Delta T/2]$ considered.

We have shown[17] that the ensemble average $\langle |S_{DC,m}(\tau)|^2 \rangle$ corresponds to the time-resolved averaged scattered intensity. The purpose of this paper is to analyze correlations like $I_p(\tau) = \langle S_{DC,m}(\tau) S^*_{DC,m+p}(\tau) \rangle$, where we assume that the field fluctuations are stationary. Using the fact that, thanks to the randomness in the field phase, $\langle \mathcal{A}\mathcal{A} \rangle = \langle \mathcal{A}^*\mathcal{A}^* \rangle = 0$ for the scattered field, a straightforward calculation leads, for even values of *p*, to:

$$I_p(\tau) \propto \int \mathfrak{R}_e\{\Phi(\tau',\omega,t_2-t_1+p\Delta T/2)G^*(\tau',\omega,t_1,t_2)\}\times\cdots$$
$$\cdots\text{Ref}(t_1,\tau)\text{Ref}^*(t_2,\tau)\times dt_1 dt_2 d\tau' d\omega \tag{5}$$

where, if the Wigner transform WT is defined by[22]:

$$WT\{f(\tau_1,\tau_2)\}(\tau,\omega) = \int f(\tau+\frac{T}{2},\tau-\frac{T}{2})\exp[-i\omega T]dT ,$$

$$\Phi(\tau,\omega,t_2-t_1) = WT\{\mathcal{A}(\tau_1,t_1)\mathcal{A}^*(\tau_2,t_2)\}$$

and

$$G(\tau,\omega,t_1,t_2) = WT\{\exp[i\omega(t_1)\tau_1 - i\omega(t_2)\tau_2]\},$$

and where the integration interval is set to be $[m\Delta T/2, (m+1)\Delta T/2]$ for both $t_1$ and $t_2$, leading to the presence of the $p\Delta T/2$ term. It can be readily shown that $G(\tau,\omega,t_1,t_2) \propto \delta[\omega - (\omega(t_1) + \omega(t_2))/2]$, so that $\omega$ in (5) lies in the modulation range.

Let us now examine the field correlation function $\Phi(\tau,\omega,t)$. The scattered field can be written as a sum of the different paths contributions, each with the form $\tilde{s}_\kappa(\omega) = \tilde{s}_i(\omega) t_\kappa \exp(-i\omega \frac{\ell_\kappa}{c})$, where $\kappa$ is indexing the path, $\ell_\kappa$ is the path length, $\tilde{s}_i$ is the amplitude of the source, $t_\kappa$ is an amplitude transmission coefficient and $c$ the speed of the light in the medium. The Fourier transform of $\Phi(\tau,\omega,t)$ with respect to $\tau$, that is $\tilde{\Phi}(\Omega,\omega,t)$, can be evaluated using the ladder diagram approximation[7], leading to:

$$\tilde{\Phi}(\Omega,\omega,t) = \tilde{s}_i(\omega+\frac{\Omega}{2})\tilde{s}_i^*(\omega-\frac{\Omega}{2}) < \sum_\kappa \mathcal{T}_\kappa \exp[i\Delta\phi_\kappa(t)] \exp[-i\Omega\frac{\ell_\kappa(0)}{c}] > \quad (6)$$

Where
$$\Delta\phi_\kappa(t) = \left(\omega - \frac{\Omega}{2}\right)\frac{\ell_\kappa(t) - \ell_\kappa(0)}{c}$$

As the photon transit time $\tau$ is of the order of hundreds of picoseconds, $\Omega$ is in the tens gigahertz range, and is negligible compared to the radian frequency ω of the light source. Furthermore we make the basic assumption[17] that the optical coefficients do not significantly vary with the source wavelength, and we consequently omit the wavelength dependence of the light speed $c$ and of the energy transmission coefficient $\mathcal{T}_\kappa$. The average of the Doppler shift term in (6) can be separated for the paths distribution[7] if we are considering paths with a given scattering number $n$:

$$\widetilde{\Phi}(\Omega,\omega,t) = \sum_n g_1(t,n) \times E_i f(\omega) < \sum_{\kappa, n(\kappa)=n} \mathcal{T}_\kappa \exp[-i\Omega \frac{\ell_\kappa(0)}{c}] > \qquad (7)$$

where $g_1(t,n) = <\exp[i\Delta\phi_\kappa(t)]>_{n(\kappa)=n}$ is the well-known autocorrelation function[9] restricted to trajectories with $n$ scattering events, and where we have set $|\widetilde{s}_i(\omega)|^2 = E_i f(\omega)$, $f(\omega)$ being the normalized spectrum profile (which satisfies $\int f(\omega) d\omega = 2\pi$). Those different approximations allow to clearly separate in (7) the contributions of $\omega$, $\Omega$ and $t$. We obviously have $\int \Phi(\tau,\omega,0) d\omega = 2\pi < |s(\tau)|^2 >$, which corresponds to the average scattered intensity $I(\tau)$. If we introduce the average scattered intensity restricted to paths with $n$ scattering events $I^{(n)}(\tau)$, or more precisely its Fourier transform $\widetilde{I}^{(n)}(\Omega)$, we can therefore identify (7) as:

$$\widetilde{\Phi}(\Omega,\omega,t) \propto f(\omega) \sum_n g_1(t,n) \widetilde{I}^{(n)}(\Omega)$$

Using the fact that, in the diffusing regime, a path length $\ell = c\tau$ corresponds very accurately to $n = \ell/l = \mu_s c\tau$ scattering events[7], where $l$ is the scattering mean free path and $\mu_s = 1/l$ is the scattering coefficient, we can then write in the real space:

$$\Phi(\tau,\omega,t) \propto f(\omega) g_1(t, \mu_s c\tau) I(\tau) \qquad (8)$$

where we recall that $f(\omega)$ can be considered as a constant when $\omega$ is in the modulation range.

We can now insert (8) in (5) in order to derive an expression that allows a quantitative analysis of the influence of the correlation function on the time resolution of our setup. The wavelength modulation frequency must indeed be high enough in order to freeze the speckle pattern fluctuations during half a modulation period. We have already shown[18] that a wavelength modulation frequency of *300Hz* is high enough to work with viscous fluids or with some biological tissues. We will therefore make the additional assumption that the correlation function is almost constant during half a modulation period and

$$g_1(t_2 - t_1 + p\Delta T/2, \mu_s c\tau) \approx g_1(p\Delta T/2, \mu_s c\tau) \qquad (9)$$

since the integration interval, for both $t_1$ and $t_2$, is set to be $[m\Delta T/2, (m+1)\Delta T/2]$

The expression (5) can now be easily calculated:

$$I_p(\tau) \propto \int I(\tau') g_1(p\Delta T/2, \mu_s c\tau') [\Pi(\tau - \tau') + \Pi(\tau + \tau')] d\tau' \qquad (10)$$

where $\Pi(\tau) \propto \frac{1}{\tau^4} J_2^2(\Delta\Omega \tau)$ is a gate function and where $J_2$ is the second order Bessel function.

The time resolution of our setup is therefore determined by the modulation depth $\Delta\Omega$ of the wavelength scan: a peak to peak frequency modulation of 30GHz thus corresponds to a 45ps time resolution [17,18], which is reasonably small compared to the typical times of flight we will observe. One should furthermore note that, for $\tau > 0$, the term $\Pi(\tau+\tau')$ can be neglected in (10), and we have approximately:

$$I_p(\tau) \propto I(\tau) g_1(p\Delta T/2, \mu_s c\tau) \qquad (11)$$

For $p=0$, $g_1 = 1$ and $I_0(\tau) = I(\tau)$ : we find the time-resolved average intensity, as previously shown[17,18]. The very new point is that the ratio $I_p/I_0$ is directly related to the time-resolved correlation function, which acts as an additional attenuation factor and can be written:

$$g_1(t, \mu_s c\tau) = \exp[-\mu_f(t) c\tau] \qquad (12)$$

This simply comes from the fact that the contributions of the different scattering events to the Doppler shift are independent. Thanks to time-resolution, the effective absorption coefficient $\mu_f$ can be straightforwardly deduced from (12). This is not so easy in standard DWS, where a more complex inverse problem must be performed, taking into account boundary conditions and other optical coefficients.

It is well known[10] that $\mu_f$ is directly linked to the microscopic movements in the medium. For the Brownian motion, we have:

$$\mu_f(t) = 2\mu'_s |t|/t_0 \qquad (13)$$

where $\mu'_s$ is the reduced scattering coefficient[23] and $t_0 = 1/k^2 D_B$ is the characteristic diffusion time. $k$ is the wave number and $D_B$ is the diffusion constant of the Brownian motion. For spherical scatterers of radius $a$ and at temperature $T$:

$$D_B = \frac{6\pi\eta a}{k_B T} \qquad (14)$$

We are now going to show that our method can be used experimentally in the diffusion regime to perform time-resolved measurements of the correlation function since we can restrict the measurement to a selected length of the photon path. Moreover, with our setup, these selected photon path lengths can exceed several hundreds transport mean free paths.

Our experimental setup was described in a previous paper [18] and is shown in fig. 1. Let us recall its main features: The wavelength modulated source is a Littman extended cavity laser diode (TEC-500-780-30) emitting 7mW at λ=780nm, with a line width of about 1 MHz. This source allows a modehop free modulation of the laser frequency of about 30GHz at a modulation frequency $f = 300 Hz$. In order to cancel parasitic signals from the reference beam we use an optical isolator, a low reference signal (*1μW* in the following), and a balanced detection. An acousto-optic modulator (AOM in fig. 1), placed in the signal arm, plays the role of an optical shutter in order to perform a real-time background subtraction: the signal is acquired during one modulation period, and the background is measured during the next period when the shutter is closed. The periodicity of the acquisition process therefore corresponds to two modulation periods. Graded-index multimode fibers where used for both the signal and the reference arms.

Concerning the sample, we used a phantom made with calibrated polystyrene microspheres (diameter *520±37* nm, refractive index *1.580* at *780nm*) in suspension in a viscous liquid in order to freeze the field fluctuations in the scale of the modulation period presently used (we estimate that a modulation frequency f = 10kHz is needed for water). Since polystyrene microspheres are usually prepared in water suspensions, we used suspensions in glycerol which is miscible with water. Glycerol has a very weak absorption coefficient in the near infrared range[24]. Its viscosity $\eta$ is high and depends on both temperature and water concentration[25]. The refractive index of pure glycerol is *n=1.472*, and depends also on water concentration[26]. In the experiment considered in this paper, the reduced scattering coefficient was set to $\mu'_s$ = *17.5cm$^{-1}$*, which corresponds to a water concentration of 7.6%, a refractive index of *1.461* and a viscosity $\eta$ =*275±10mPa.s* at 22.5±0.5°C.

The measurements were performed with a reflectance geometry. The emission and detection fibers are perpendicular to the sample surface, just in contact with the liquid, with a source-detector separation *r=1cm*. A black optical shield is placed between the tips of these fibers in order to avoid cross-talking. The ensemble averaging was performed over 90 000 modulation periods. The reduced scattering coefficient of the sample was experimentally checked from the time-resolved reflectance function measured by our system and fitted by Monte Carlo simulations[27]: we obtained $\mu'_s$ = *17.6±0.5cm$^{-1}$*, in good accordance with the awaited value.

We have calculated the ratio $I_p/I_0$ for different $p$ values. In fact, as the periodicity of the acquisition procedure corresponds to 2 modulation periods, that is to 4 times T/2, we set $p=4p'$. Fig. 2 presents the results obtained for p' varying from 1 to 5. For p'=1 a signal has been obtained up to $\tau$=1000ps, which corresponds to more than 350 transport mean free paths, and to about 2000 scattering events. As predicted, the $ln(I_p/I_0)$ curves exhibit a linear behavior with the photons time of flight $\tau$. The values of the effective absorption coefficient $\mu_f(t_{p'})$ obtained from the linear curves fits (thin curves in fig.2) are plotted in fig. 3 as a function of $t_{p'} = 2p'\Delta T$.

The results shown in fig. 3 can be fitted by $\mu_f(t) = \alpha t$. The fit, limited to the 3 first values of p' as the uncertainty is quite high for p'=4 and 5, gives $\alpha$=15.2±0.2$cm^{-1}s^{-1}$. For the Brownian motion $\alpha = 2\mu'_s/t_0$ and we obtain a value of the characteristic diffusion time $t_0$ = 2.3±0.1 s, to be compared with the awaited value at 22.5°C: $t_0$ = 2.4±0.2 s.

In conclusion, we have reported the first experimental evidence for time-resolved diffusing wave spectroscopy with photon path lengths that overpass 300 transport mean free paths. Such measurements can simplify the analysis of correlation measurements. The major advantage of this method is the possibility to get all benefits of time resolution for sensing deep layers in biological tissue, or for improving depth resolution in transillumination. The next step will be to increase the modulation frequency (up to 10kHz[28]) in order to work with biological media.

* Corresponding author email address: tualle@galilee.univ-paris13.fr

**Figure captions**

**Fig. 1:** experimental setup

**Fig. 2:** Experimental curves $\ln(I_{p'}/I_0)$ versus time $\tau$, for different values of the parameter $p'$ ($t=2p'\Delta T$), with their linear fit.

**Fig. 3:** Values of $\mu_f$ obtained from the fits of fig. 2 for the different values of p'. The 3 first values (for $p'=1,2,3$) allows a determination of the slope $\alpha$.

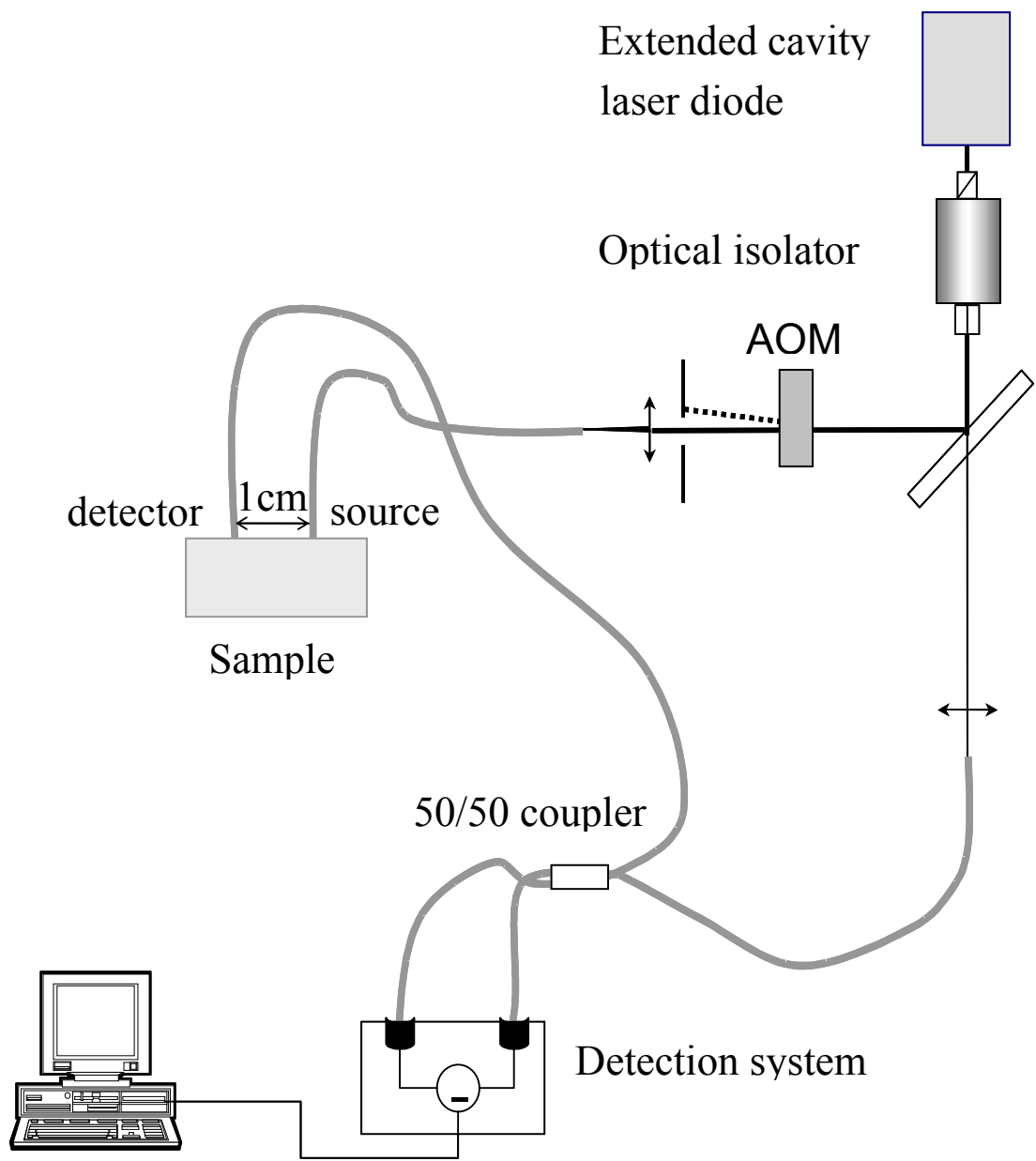

fig. 1 :

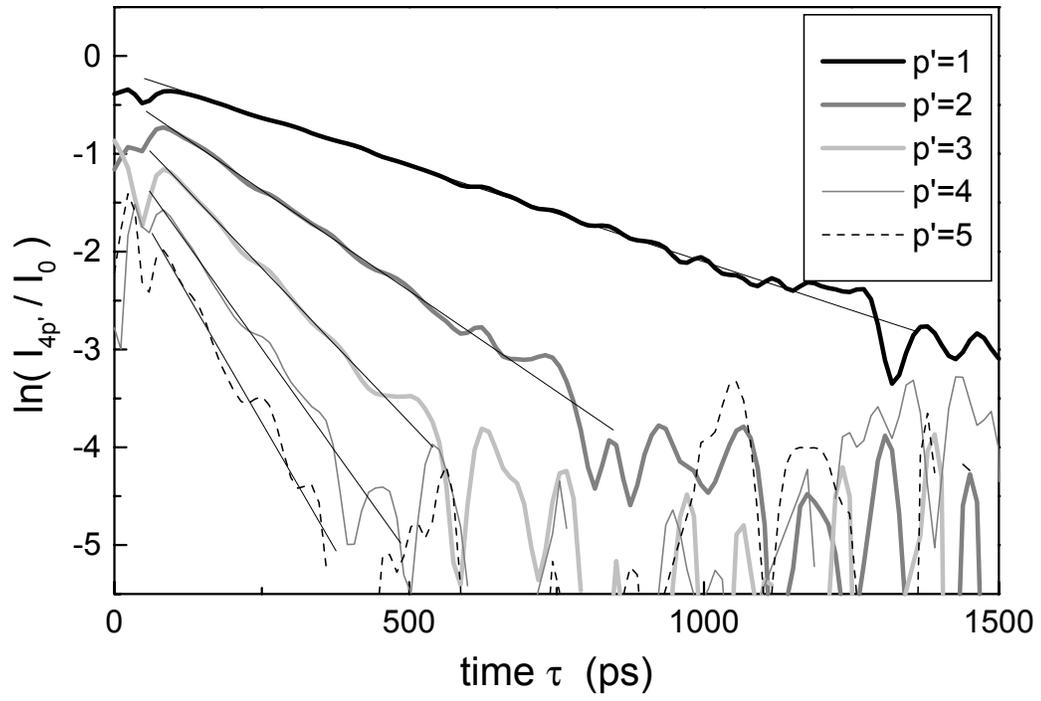

fig. 2 :

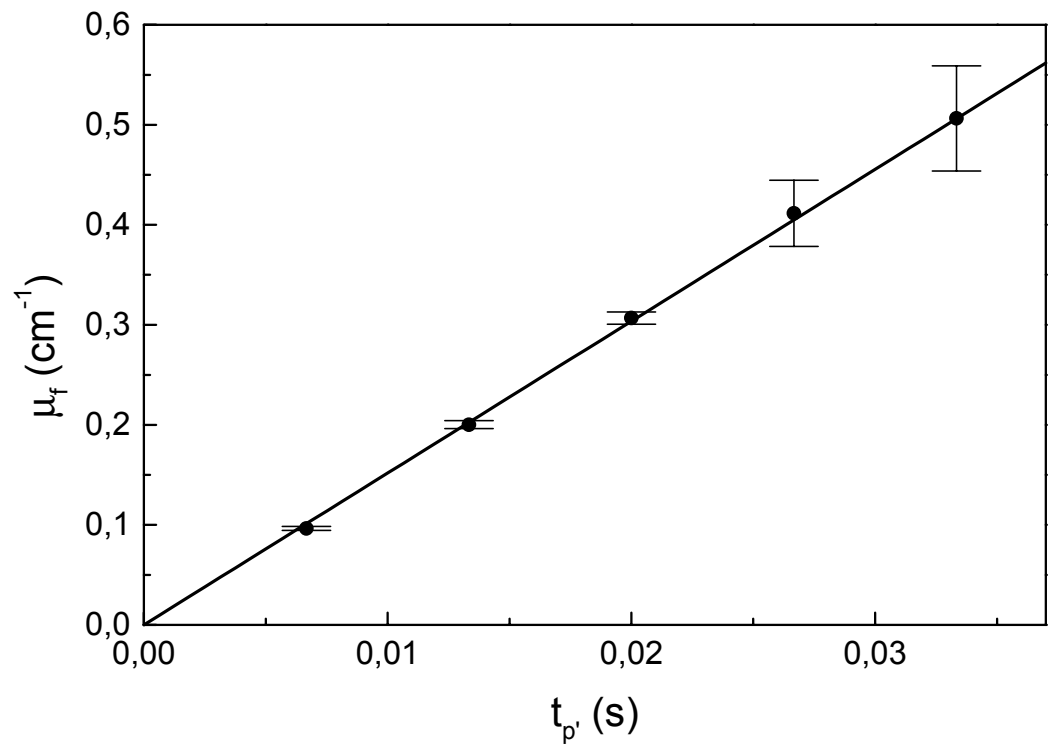

fig. 3 :